\documentclass[twocolumn,showpacs,preprintnumbers,amsmath,amssymb]{revtex4}

\usepackage{graphicx}
\usepackage{dcolumn}
\usepackage{bm}

\begin{document}
\title{Parameter-tuning Networks: Experiments and Active Walk Model}

\author{Xiao-Pu Han$^{1, c}$}\
\email{hxp@mail.ustc.edu.cn}
\author{Chun-Dong Hu$^{2}$}\
\author{Zhi-Min Liu$^{2}$}\
\author{Bing-Hong Wang$^{1,3}$}
\email{bhwang@ustc.edu.cn}

\affiliation{$^{1}$ Department of Modern Physics, University of
Science and Technology of China, Hefei, 230026, China\\$^{2}$
Institute of Plasma Physics, Chinese Academy of Sciences, Hefei,
230031, China\\  $^{3}$ Shanghai Academy of System Science and
University of Shanghai for Science and Technology, Shanghai, 200093
China}

\date{\today}

\begin{abstract}
The tuning process of a large apparatus of many components could be
represented and quantified by constructing parameter-tuning
networks. The experimental tuning of the ion source of the neutral
beam injector of HT-7 Tokamak is presented as an example.
Stretched-exponential cumulative degree distributions are found in
the parameter-tuning networks. An active walk model with eight
walkers is constructed. Each active walker is a particle moving with
friction in an energy landscape; the landscape is modified by the
collective action of all the walkers. Numerical simulations show
that the parameter-tuning networks generated by the model also give
stretched exponential functions, in good agreement with experiments.
Our methods provide a new way and a new insight to understand the
action of humans in the parameter-tuning of experimental processes,
is helpful for experimental research and other optimization
problems.
\end{abstract}
\maketitle

Large apparatuses pay more and more important role in the
experimental research of modern physics, such as the high-energy
accelerators and controlling nuclear fusion devices. A large
apparatus made up of many components, each component needs be tested
and tuned separately and then collectively, before experiments using
the apparatus as a whole can be conducted. And there are a large
number of parameters to be tuned. The parameter tuning processes are
often complicated and onerous, strongly depend on the the experience
of experimenter. How can we model the adjustment process of these
parameters and make some sense out of it?

For example, in a Tokamak in plasma physics, in the ion
source segment of the neutral-beam injector system alone, there are
eight major parameters to be adjusted. The experimenter sets the
parameter values, turns on the equipment, and measures the outcome
of a certain quantity $Q$. If the $Q$ obtained does not meet the mark
$Q_{0}$, say, the whole process is repeated, with a new set of
parameters. The adjustment process stops when $Q$ is equal or very
close to $Q_{0}$. The simultaneous adjustment of a large number of
parameters is a complicated process, which is based on the feedback
from previous $Q$ values obtained, the experimenter's experience, and
the limitation of the hardware that control the parameters.

To quantify this complicated process, a parameter-tuning network can
be constructed as follows. Assume that $N$ parameters $u_{i} (i =
1,2,\cdots, N)$ are involved. The experimenter's serial adjustment
of $\emph{\textbf{u}}$ $[\equiv (u_{1},\cdots, u_{N})]$ can be
represented by a sequence of connected dots in the $N$-dimensional
$\emph{\textbf{u}}$ space. For each choice of $\emph{\textbf{u}}$,
$Q$ is measured, so that $Q = Q(\emph{\textbf{u}})$ but the
functional form is unknown due to the complexity of the equipment,
which is like a black box.

The sequence of $\emph{\textbf{u}}$ dots is then projected onto the
$(u_{1}, u_{2})$ plane, say. The dots on the same vertical line
along the $u_{3}$ direction are allowed to collapse to one point,
called a node. A line connecting two nodes is called an edge
\cite{Wat1,Bar1,Bar2}. Since in real situations, the parameters
selected by the experimenter in different adjustments may partially
overlap with each other, there may be more than one edge connecting
two nodes. These edges are directed. For simplicity, we make the
approximation of collapsing all the edges between two nodes that
have the same direction as one single edge with the same direction.
Consequently, between two nodes, there are at most two edges with
opposite directions, resulting in a \emph{directed} parameter-tuning
network. A \emph{non-directed} network is formed from the directed
one, by collapsing all the edges between any two nodes and removing
the directions.

\begin{figure}
  \includegraphics[width=8.6cm]{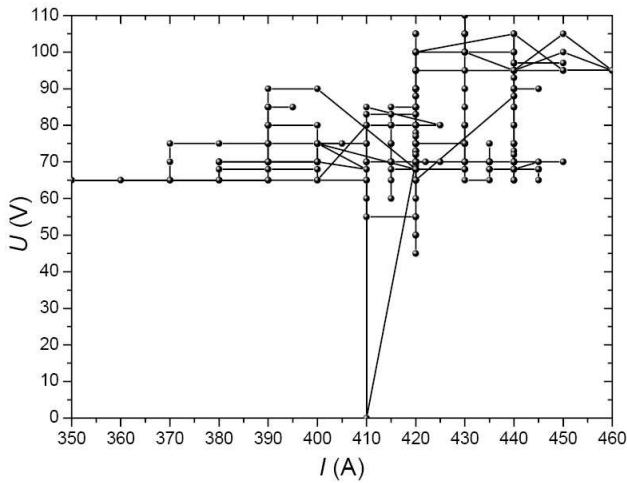}\\
  \caption{The non-directed parameter network on the (\emph{U}, \emph{I}) plane
derived from the experimental
 data, where \emph{U} is the working voltage of the cathode
  gas valve and \emph{I} the filament current.}
\end{figure}

\begin{figure}
  \includegraphics[width=8.6cm]{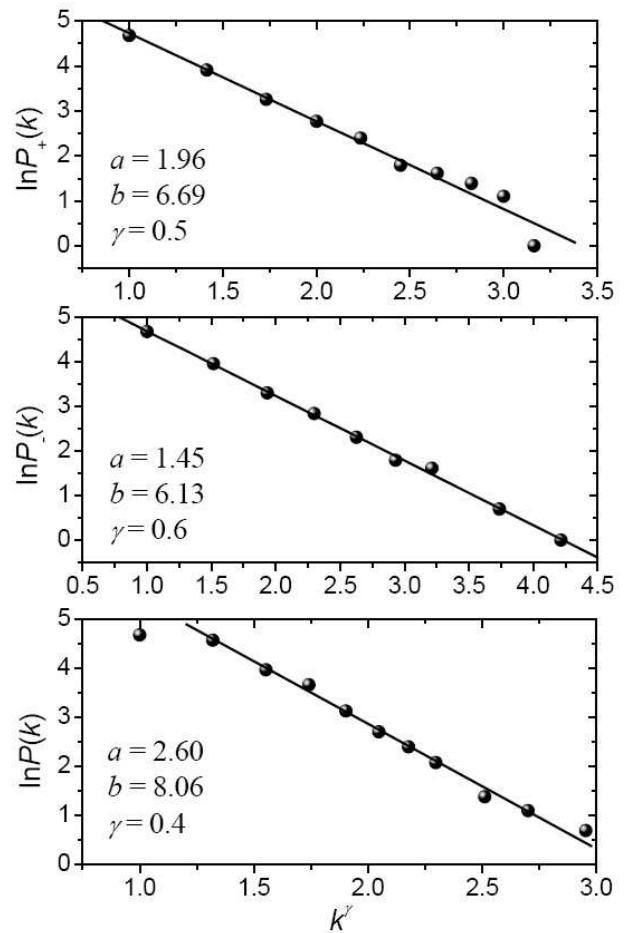}\\
  \caption{Experimental curves of $P_{+}(k)$, $P_{-}(k)$ and $P(k)$, fitted to stretched
  exponent functions. The corresponding (directed and non-directed)
  parameter-tuning networks exist on the (\emph{U}, \emph{I}) plane, where \emph{U} is the working voltage of the cathode
  gas valve and \emph{I} the filament current. Its non-directed structure has been shown in FIG. 1. The experimental data obtained from the tuning of the ion source
  in a whole experiment season, the sequence includes 789 parameter sets.}
\end{figure}

In a parameter-tuning network, there are $N_{n}$ nodes, say. For a
directed parameter-tuning network node $j (j = 1, 2, \cdots, N_{n})$, let
$N_{+}(j)$ be the number of in-going edges, $N_{-}(j)$ the number of out-going
edge.

Now count the number of $N_{+}$ that has the value $k$, and call it
$p_{+}(k)$; similarly for $p_{-}(k)$. The corresponding function in
the non-directed network case is denoted by $p(k)$. $k$ is called
the "degree" of a node. $p_{+}(k)$ is called the "in-degree
distribution", $p_{-}(k)$ the "out-degree distribution", and $p(k)$
the "degree distribution." Next define the corresponding
"culmulative degree distributions" $P_{+}(k)$, $P_{-}(k)$ and $P(k)$
by
\begin{equation}P_{+}(k) \equiv \sum_{k'=k} ^{k_{m}} p_{+}(k')\\
\end{equation}
\begin{equation}P_{-}(k) \equiv \sum_{k'=k} ^{k_{m}} p_{-}(k')\\
\end{equation}
\begin{equation}P(k) \equiv \sum_{k'=k} ^{k_{m}} p(k')\\
\end{equation}
Here $k_{m}$ is the maximum $k$ that the $p$ function is non-zero, and
all of the distribution functions are not unitary. Note that
by definition, the $P$'s are monotonic decreasing functions, while the
$p$'s may not be monotonic at all. The reason for introducing the $P$'s
is that we want to fit them to monotonic decreasing functions such
as stretched exponents or power laws.

In our experiments studying the ion source of neutral beam injector
system for HT-7 Tokamak, eight control parameters are involved $(N =
8)$. These include the filament current, magnet current, arc voltage,
cathode gas valve voltage, anode gas valve voltage, etc. Each of the
parameters has 10 to 30 discrete, adjustable setting values. The aim
of the experiment is to find out which set of parameters will give a
strong and stable discharge, measured by the arc current intensity
$Q$. Each parameter is set by a dial which can be turned left or right
between two extreme positions. When the extreme position is reached,
the experimenter has to turn the dial back, reversing the direction
of turning.

From our experimental data, the sequence of parameters in the
8-dimensional $\emph{\textbf{u}}$ space is first generated and the
corresponding directed and non-directed tuning-parameter networks
are constructed. The non-directed network is shown in FIG. 1. The
$P$'s are obtained, and fitted nicely with stretched exponential
functions (FIG. 2) such that
\begin{equation}P(k) \sim  \exp(-ak^{\gamma})
\end{equation}
or, equivalently,
\begin{equation}\ln P(k) = - ak^{\gamma} + b
\end{equation}
and similarly for $P_{+}(k)$ and $P_{-}(k)$. Such property is also
found in the parameter-tuning networks derived from the data of the
tuning experiments in other experiment season of the ion source
(each of them includes 600 to 800 parameter sets).

The experimental adjustment of the parameters is somehow correlated
by the experimenter, and the process can be modeled by an active
walk (AW) model \cite{Lam2}. In our AW model here, $N = 8$; each
$u_{i}$ has adjustable values of $1, 2,\cdots, 20$, say. The
adjustment of the $i$th parameter is represented by the movement of
an active walker on a 1D landscape potential $V_{i}(x_{i})$. The
allowable $x_{i}$ for the $i$th walker are the integers $(1, 2,
3,\cdots, 100)$, the same for all $i$. These 100 numbers are mapped
to the $u_{i}$, such that if the $i$th walker ends in the region
$[1, 5]$ on the $x_{i}$ axis, $u_{i}$ will assume the value of 1.
Similarly, the region of $[6, 10]$ is mapped to $u_{i} = 2$, etc.
This mapping can be attained by
\begin{equation}u_{i}(t)= Int[(X_{i}(t)-1)/5] +1
\end{equation}
where $X_{i}(t)$ is the position of the $i$th walker on the $x_{i}$
axis at time $t$, and $Int$ means taking the integral part of the
number. This mapping of $x_{i}$ to $u_{i}$ has the effect of making
two consecutive sets of adjusted parameters more likely to partially
overlap with each other, and ensures the occurrence of smooth
$V_{i}$ vs $x_{i}$ curves.

\begin{figure}
  \includegraphics[width=8.6cm]{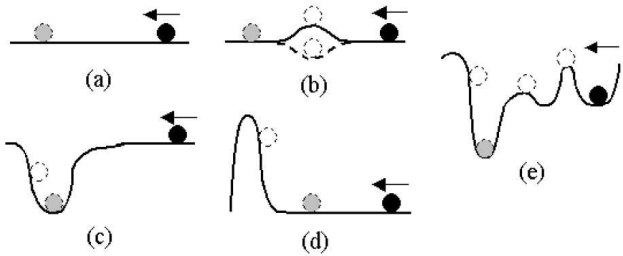}\\
  \caption{Sketch of possible subwalks of a particle. The solid dot,
  gray dot and open circle represent the initial position, finial
  position and some intermediate position of the particle, respectively;
  the arrow indicates the initial direction of the particle. (a) The
  particle moves and stops on a plateau. (b) The particle gets over a
  low potential barrier or a shallow well. (c) The particle stops in a deep
  well. (d) The particle rebounds on a high potential barrier. (e) The
  particle moves on a general landscape, and ends up in a deep well after many steps.}
\end{figure}

In the simulations, all $V_{i}$ start flat at time $t = 0$, and are updated
simultaneously at each time step $t (=1, 2,\cdots)$, according to a rule
to be specified below. But between two consecutive time steps, each
walker moves a few steps in a subwalk in its own $V_{i}$, independent of
other walkers. In a subwalk, the walker does not change $V_{i}$. The
subwalk is like a particle rolling on a landscape with friction,
with the following rules (with the subscript $i$ removed for the sake
of clarity). The subwalk time is labeled by $\tau (= 0, 1, 2,\cdots)$.\\
(i). At time $t = 0$, the particle is arbitrarily placed on the $x$ axis.\\
(ii). At time $t$, the particle is given energy $K_{0}$ at $\tau  = 0$. ($K_{0}$ is a
parameter fixed in the model.)\\
(iii). At subwalk time $\tau$, the particle moves left or right with
equal probability. However, after each move, the particle loses
energy $\varepsilon$, and the energy difference $V(x(\tau))-V(x(\tau + 1))$
which could be positive or negative is added to its
energy. Consequently, at time $\tau$, the particle already moves
$\tau$ steps, and its energy becomes
\begin{equation}K(\tau) = K_{0}+ V_{0}-V(x(\tau))- \tau\varepsilon
\end{equation}
where $V_{0} \equiv V(x(0))$, the potential at the initial position of the
particle at $\tau = 0$.\\
(iv). The particle can get over a potential
barrier in $V$ that is lower than $K$, and can rebound if the barrier is
higher than $K$.\\
(v). When the particle reaches the left boundary $(x =
1)$ or the right boundary $(x = 100)$, it reverses direction and
continues its subwalk.\\
(vi). The particle stops when it walks into a
potential well and cannot get out with its energy $K$, or exhausts its
energy on a plateau.\\

These possibilities are sketched in FIG. 3. The reflecting boundary
condition in item (v) corresponds to the real experimental situation
where a dial with a limited range is used.
\begin{figure}
  \includegraphics[width=8.6cm]{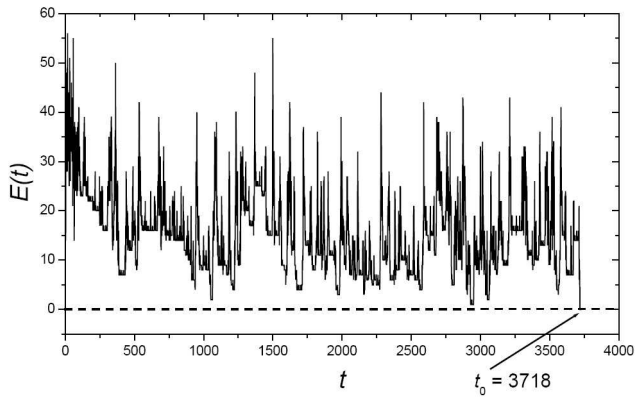}\\
  \caption{Dependence of the error function on time, from the eight-parameter AW model simulation.}
\end{figure}

\begin{figure}
  \includegraphics[width=8.6cm]{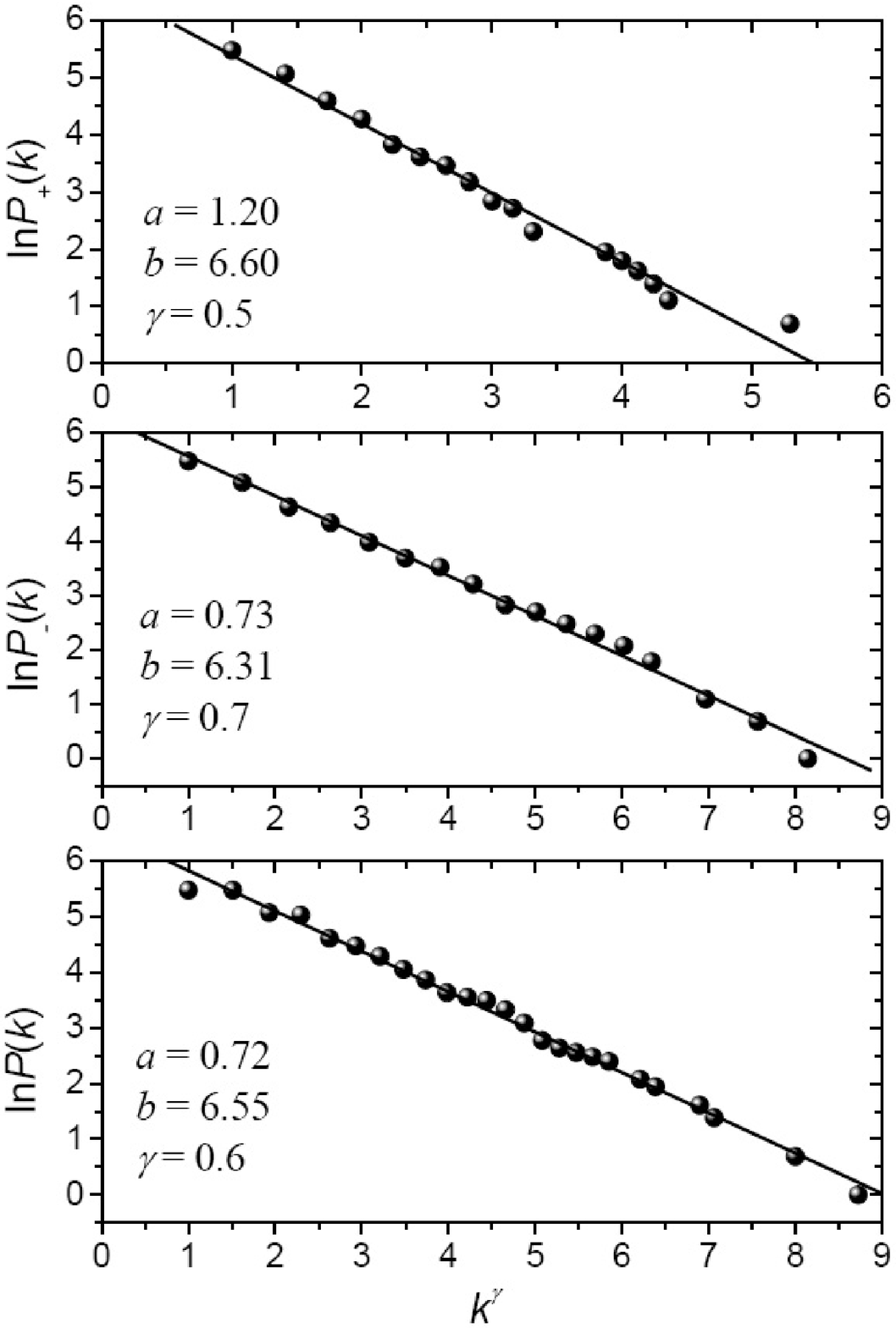}\\
  \caption{Numerical curves of $P_{+}(k)$, $P_{-}(k)$ and $P(k)$ obtained from the
  parameter-tuning networks from a eight-parameter AW model
  simulation, fitted to stretched exponent functions.}
\end{figure}

After all $N$ particles stop, the time clock increases from $t$ to
$t + 1$; the particle's position at the end of its subwalk is taken
to be $X(t + 1)$, which is the starting position of the subwalk at
time $t + 1$. (The subwalks of the particles may not stop after the
same number of subwalk steps; those stop first will sit there and
wait for the last particle to stop.) The landscape $V_{i}(x_{i})$ is
updated by the landscaping rule:
\begin{equation}
    V_{i}(x_{i};t+1)=\left\{
    \begin{array}{cc}
V_{i}(x_{i};t)+W(x_{i}-X_{i}(t+1)),\\
E(t)\geq E_{a}(t);\\
V_{i}(x_{i};t)-W(x_{i}-X_{i}(t+1)),\\
E(t)<E_{a}(t).
    \end{array}
    \right.
\end{equation}
where $t = 0, 1, 2,\cdots$. In our numerical simulations below, the
($i$-independent) landscaping function $W$ is given by $W(0) = 4$,
$W(\pm1) = 3$, $W(\pm2) = 1$, and $W = 0$ otherwise. In Eq. (8), the
error function $E(t)$ is assumed to be
\begin{equation}E(t) = |Q(\emph{\textbf{u}}(t)) - Q_{0}|
\end{equation}
where $\emph{\textbf{u}}(t)$ is obtained from Eq. (6); $E_{a}(t)$ is
the average error counting the last $n$ time steps, given by
\begin{equation}E_{a}(t)=\frac{1}{n}\sum ^{t}_{t'=t-n+1}E(t')
\end{equation}
In computer simulations, we pretend that we know $Q_{0}$ and the
$Q(\emph{\textbf{u}})$ function. In reality, the former is known to
the experimenter; the latter can be roughly inferred from
experimental data. Note that while the subwalks of the particles are
independent of each other, the updating of their landscapes is
affected by their collective effort through $Q(\emph{\textbf{u}})$
in $E(t)$.

The eight-parameter AW model is simulated with $t$ going from 0, and
the simulation is stopped when the optimal parameter set is gotten.
The function $Q$ is assumed to be
\begin{equation}Q(\emph{\textbf{u}})=|u_{1}-10|+|u_{2}-10| +\cdots+|u_{8}-10|
\end{equation}
The parameters used are: $K_{0} = 23$, $\varepsilon = 1$, $n = 5$
and $Q_{0} = 0$. This means that $\emph{\textbf{u}} = (10, 10,
\cdots, 10)$ is the one and only optimal parameter set.

The error function $E(t)$ is displayed in FIG. 4, which shows that
the optimal $\emph{\textbf{u}}$ is first obtained at $t_{0} = 3718$.
The 700 $\emph{\textbf{u}}(t)$ dots (with $300 < t \leq 1000$, such
number of $\emph{\textbf{u}}(t)$ is similar with the experimental
data, and the initialization process when $t<300$ is ignored) in the
$\emph{\textbf{u}}$ space are projected onto the $(u_{1}, u_{2})$
plane to obtain the directed and non-directed networks. The
corresponding cumulative degree distributions are given in FIG. 5.
Good stretched-exponential fits are obtained, in agreement with the
experimental findings in FIG. 2.

In the AW model, the subwalks kind of mimicking the action of the
experimenter--we think that is why the AW model and the experiments
give similar $P$ functions. Human dynamics \cite{Bar3,Vqz9} have
been studied in other situations by different models. Our results
imply that our network-based statistics and the AW model is helpful
in the understanding of human action in the experimental processes
and very likely in many other optimization projects. The stretched
exponent property of the parameter-tuning network imply the human
action in the optimization process could be having some quantitative
principles, which is important for the experimental studies.
Stretched exponent distributions or relations exist widely in many
other systems \cite{XB4,Hol5,Sor1,Str1}. Since AW-like dynamics is
very common in nature \cite{Lam2}, our model may reveal a new
mechanism of stretched exponent distribution. By modifying the
subwalks and tuning their parameters one may shorten $t_{0}$, the
shortest time to find the optimal parameters, which would be of
interest to the experimental researchers. Finally, the AW model
mimics an optimization process in real experiment, could be useful
in the optimization projects in many other systems.

We thank Lui Lam, Pin-Qun Jiang, Da-Ren He and Tao Zhou for useful
discussions, and Sheng Liu, Shi-Hua Song, Jun Li and Yuan-Lai Xie
for experimental help. This work is partially supported by China 973
Program (Grant No. 2006CB705500) and NSF China (Grant Nos. 60744003,
10635040, 10532060, 10472116, and 10575105).

\end{document}